\documentclass[aps,prl,twocolumn, superscriptaddress,letterpaper]{revtex4}
\usepackage{graphicx}
\usepackage{amssymb,amsmath}
\usepackage{float}
\usepackage{bm}

\usepackage{setspace}
\usepackage{parskip}
\usepackage{xcolor}
\definecolor{midnightblue}{cmyk}{1,1,0,0.1}
\definecolor{forestgreen}{cmyk}{0.76,0,0.26,0.5}

\usepackage{epsfig}

\usepackage{hyperref}
\hypersetup{
    bookmarks=true,         % show bookmarks bar?
    unicode=false,          % non-Latin characters in Acrobat’s bookmarks
    pdftoolbar=true,        % show Acrobat’s toolbar?
    pdfmenubar=true,        % show Acrobat’s menu?
    pdffitwindow=false,     % window fit to page when opened
    pdfstartview={FitH},    % fits the width of the page to the window
    pdftitle={My title},    % title
    pdfauthor={Author},     % author
    pdfsubject={Subject},   % subject of the document
    pdfcreator={Creator},   % creator of the document
    pdfproducer={Producer}, % producer of the document
    pdfkeywords={keyword1} {key2} {key3}, % list of keywords
    pdfnewwindow=true,      % links in new window
    colorlinks=true,       % false: boxed links; true: colored links
    linkcolor=midnightblue,          % color of internal links (change box color with linkbordercolor)
    citecolor=magenta,        % color of links to bibliography
    filecolor=midnightblue,      % color of file links
    urlcolor=midnightblue,          % color of external links
}

\begin{document}

\title{Out-of-plane carrier spin in transition-metal dichalcogenides under electric current}

\author{Xiao Li}
%\email{lixiao150@utexas.edu}
\affiliation{Center for Quantum Transport and Thermal Energy Science,
School of Physics and Technology, Nanjing Normal University, Nanjing 210023, China}
\affiliation{Department of Physics, University of Texas at Austin, Austin, TX 78712, USA}

\author{Hua Chen}
\affiliation{Department of Physics, Colorado State University, Fort Collins, CO 80523, USA}
\affiliation{School of Advanced Materials Discovery, Colorado State University, Fort Collins, CO 80523, USA}

\author{Qian Niu}
%\email{niu@physics.utexas.edu}
\affiliation{Department of Physics, University of Texas at Austin, Austin, TX 78712, USA}
%\affiliation{International Center for Quantum Materials, School of Physics, Peking University, Beijing 100871, P. R. China }
%\affiliation{Collaborative Innovation Center of Quantum Matter, Beijing, P. R. China }

\begin{abstract}
In a multilayer comprising ferromagnet and heavy metal, in-plane carrier spin is induced by applied electric current owing to Rashba spin-orbit coupling, while the out-of-plane component is absent. 
We propose the out-of-plane carrier spin can emerge in ferromagnetic  transition-metal dichalcogenides monolayer, by symmetry arguments and first-principles calculations. An intrinsic spin-orbit coupling in the monolayer provides valley-contrasting Zeeman-type spin splitting for generating the vertical induced spin. The current direction can be exploited to tune the induced spin, accompanied with valley polarization.  The exotic spin accumulation paves an accessible way for perpendicular magnetization switching and electric control of valleys.

\end{abstract}

\maketitle
\textcolor{forestgreen}{\emph{\textsf{Introduction}.}}--- Non-equilibrium carrier spin and associated spin-orbit torque are induced by applied electric current in a strong spin-orbit coupled electron system, which has practical use in magnetization switching and domain-wall motion for novel spintronic devices \citep{Gambardella11}.
A typical realization of the current-induced spin reorientation is a two-dimensional heterostructure composed of ferromagnet and heavy non-magnetic layer \citep{Gambardella11, Macneill17}. The inversion asymmetry along interface normal gives arise to Rashba-type spin-orbit coupling that determines a chiral spin texture. Only in-plane induced spin emerges with the redistribution of carriers on the Fermi surface under electric current, in contrast with a vanishing normal component. However, it is highly desirable to realize out-of-plane carrier spins and magnetization reorientation.  As the magnetization is rotated from in-plane to normal direction, a variety of intriguing physical phenomena take place, e.g. topological phase transition \cite{Cai15}, semiconductor-to-metal transition \citep{Lado14} and valley splitting \citep{Qi15, Chen17}.

The emergent two-dimensional van der Waals materials % with spontaneous magnetic ordering
 provide simple and powerful platforms for the study of spin-related physics \citep{Park16, Xiao12}. Monolayer transition-metal dichalcogenides, MX$_2$ (e.g. M=V, Mo, W; X=S, Se, Te) have both strong spin-orbit coupling and inversion symmetry breaking \citep{Zhu11, Xiao12, Fuh16, Tong16}, which may lead to carrier spin under electric current.  The study of current-induced spin polarization (CISP) in MX$_2$ monolayer, though few, is fundamentally and technologically significant. Firstly,  the monolayer structure avoids complexity from the heterostructure. More importantly, the spin-orbit coupling of MX$_2$ is intrinsic and induces Zeeman-type spin splitting in the presence of in-plane mirror symmetry \citep{Zhu11, Xiao12}. While the CISP invariably relies on Rashba spin-orbit coupling in the heterostructure, the roles of the intrinsic spin-orbit coupling in CISP is not yet clear.  Moreover, there are two inequivalent valleys in the band structure of MX$_2$. The valley degree of freedom may add a new dimension to CISP, given valley-contrasting optical and electronic properties \cite{Xiao12, Cao12}.  

In this Letter, taking VSe$_2$ monolayer for example, we perform symmetry analysis and first-principles calculation to study the non-equilibrium carrier spins in transition-metal dichalcogenides,  under applied electric field. VSe$_2$ monolayer has been predicted to be a room-temperature ferromagnet \cite{Fuh16} and synthesized recently in the form of nanosheets \cite{Zhao16}.  The intrinsic spin-orbit coupling in VSe$_2$ monolayer gives rise to an out-of-plane CISP, in contrast with Rashba spin-orbit coupling.  Two valleys exhibit distinct responses to the direction of electric field, which can be used as a readily available experimental knob to tune CISP.  We also discuss the higher tunability from applied uniaxial strain, based on the same symmetry consideration. The vertical CISP provides an energy-efficient pathway towards perpendicular magnetization switching and electric control of valleys.

%Spintronic in two-dimensional materials could provide novel opportunities for future electronics, for example, efficient generation of spin current, which should enable the efficient manipulation of magnetic elements.
%The magnetic layer adjacent to two dimensional materials. Current-induced spin accumulation through the Rashba effect in the composites of magnetic metal/heavy metal  bilayer
%Current-induced spin orbit torque \citep{Gambardella11}. \citep{Culcer05,Garate09,Li16}
%Monolayer transition-metal dichalcogenides, such as MX$_2$ (M=Mo, W; X=S, Se, Te) \citep{Xiao12, Cao12} 
%Rashba-type spin orbit coupling is a momentum-dependent splitting of spin bands in two-dimensional condensed matter systems (heterostructures and surface states) similar to the splitting of particles and anti-particles in the Dirac Hamiltonian. 

\begin{figure}[h!]
\centering
\includegraphics[width=8cm]{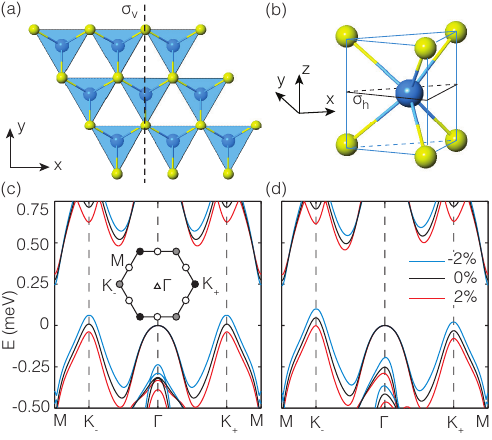}
\caption{ Atomic and band structures of VSe$_2$ monolayer. (a) Top and (b) side views of the 2H phase. Blue and yellow spheres stand for V and Se atoms, respectively. $\sigma_v$ and $\sigma_h$ are two mirror planes. (c) and (d) Band structures with the magnetization along $x$ and $z$ directions, respectively. Blue, black and red bands respectively correspond to  -2\%, 0\% and 2\% uniaxial strain along $y$ direction. The Brillouin zone and its high-symmetry points are given in the inset of (c).}
\label{fig:1}
\end{figure}

\textcolor{forestgreen}{\emph{\textsf{Symmetry analysis}.}}--- 
 Figs. \ref{fig:1}(a) and (b) show the atomic structure of VSe$_2$ monolayer in the 2H phase, with a $D_{3d}$ point group. It has a honeycomb lattice, of which two sublattices are respectively occupied by vanadium and selenium. Each vanadium atom lies in the center of a trigonal prismatic cage of six selenium atoms. The honeycomb lattice therefore has three atomic layers in a Se-V-Se stack.  The inversion symmetry is absent, which allows of non-vanishing CISP \citep{Zelezny14}.  Without taking into account the magnetization, there are two types of mirror planes. The vanadium layer is a mirror in the two-dimensional ($x$-$y$) plane of the monolayer, as denoted by $\sigma_h$  in Fig. \ref{fig:1}(b).  The other type of mirrors consists of three planes, which are perpendicular to the two-dimensional sheet and through V-Se bonds. One of them lies in $y$-$z$ plane, as denoted by $\sigma_v$. The others are respectively obtained by rotating  $\sigma_v$ through 2$\pi$/3 and 4$\pi$/3 around $z$ axis, in the presence of three-fold rotational symmetry $C_3$. 

As the symmetry argument provides an effective tool for predicting possible CISP, a symmetry analysis is firstly performed based on the above structure,  as shown in Table I.  We focus on CISP from the redistribution of carriers under the actions of electric field and disorder, which is dominant over Berry phase contributions \citep{Kurebayashi14, Zelezny14, Zelezny17, Xiong17}.  The induced spin polarization, $\mathcal{\delta} \bm s$, can be described by the relation,
\begin{equation}
\mathcal{\delta} \bm s = \tau \chi \bm E .
\label{ef}
\end{equation}
Here, $\bm E$ is applied in-plane electric field. $\tau$ is the relaxation time and $\chi$ is a material-specific coefficient. Given that the two-dimensional sheet of VSe$_2$ is an easy plane of ferromagnetization \citep{Fuh16}, we first consider a high-symmetric in-plane magnetization ($\bm {M}$) along $x$ direction. For the magnetized monolayer, the symmetry $\sigma_v$ is still present, while the other vertical mirrors and $C_3$ are broken. The joint transformation of $\sigma_h$ and time reversal $\mathcal{T}$, $\sigma_h\mathcal{T}$, becomes a new symmetry operator. Both $\sigma_v$ and $\sigma_h\mathcal{T}$ determine non-vanishing $\mathcal{\delta} \bm s$. When $\bm E \parallel x$, $\bm E$ changes its sign but $\tau$ keeps unchanged under $\sigma_v$. According to Eq. \ref{ef}, a nonzero $\chi$ requires that $\mathcal{\delta} \bm s$ changes sign, leading to $\mathcal{\delta} \bm s \parallel \sigma_v$. Under $\sigma_h\mathcal{T}$, $\bm E$ does not change but $\tau$ becomes $-\tau$. The required sign reversal of $\mathcal{\delta} \bm s$ indicates $\mathcal{\delta} \bm s \parallel z$. Therefore, for $\bm {M}\parallel x$ and $\bm E\parallel x$, only vertical component of $\mathcal{\delta}\bm s$ is allowed by considering the two mirror-related symmetries.  When $\bm E$ is reoriented along the $y$ direction, $\mathcal{\delta} \bm s$ has to keep unchanged under $\sigma_v$ and change sign under $\sigma_h\mathcal{T}$ to realize a nonzero $\chi$. Since two conditions are not satisfied simaltaneously, both $\chi$ and $\mathcal{\delta} \bm s$ are vanishing. In the same manner, a magnetization along $y$ direction is also discussed, where  $\sigma_v\mathcal{T}$ and $\sigma_h\mathcal{T}$ are symmetry operators. 

By summarizing these combinations of $\bm E$ and $\bm M$ in Table I, it is seen that no matter $\bm M \parallel x$ or $\bm M \parallel y$, $\bm E \parallel x$ has a symmetry-allowed $\mathcal{\delta} \bm s _z$, while $\bm E \parallel y$ leads to $\mathcal{\delta} \bm s =0$. Therefore, an out-of-plane CISP may emerge in VSe$_2$ monolayer and it is tunable by rotating in-plane electric field, in contrast with well-known Rashba systems. Besides the mirror-related symmetries, it is noted that the $C_3$ symmetry breaking is also important for generating $\mathcal{\delta} \bm s_z$. Since $\bm E$ is rotated by 2$\pi$/3 and $\mathcal{\delta} \bm s_z$ is unchanged under $C_3$ operation, the distinct behaviors leave Eq. \ref{ef} unsatisfied and correspondingly rule out $\mathcal{\delta} \bm s_z$ in the presence of $C_3$.  $\mathcal{\delta} \bm s$ is therefore absent in VSe$_2$ magnetized vertically and non-magnetic MX$_2$ (e.g. MoS$_2$) that have $C_3$ symmetry. 
 
\begin{table}
\caption{Current-induced spin polarization under symmetry operations, with four combinations of the electric field and magnetization considered. 
The corresponding symmetries are given in the second row.  When applying symmetry operations, the sign changes of $\bm E$, $\tau$ and $\mathcal{\delta} \bm s$  in the third to fifth rows determine non-vanishing $\mathcal{\delta} \bm s$ in the sixth row. }

\begin{center}
{\small
\scalebox{0.97}[0.97]{%
\begin{tabular}{p{10pt}|p{26pt}|p{26pt}|p{26pt}|p{26pt}|p{26pt}|p{26pt}|p{26pt}|p{26pt}}
\hline 
\hline
 & \multicolumn{2}{|c|}{$\bm {M}\parallel x$,  $\bm {E}\parallel x$}  & \multicolumn{2}{|c|}{$\bm {M} \parallel x$, $\bm {E}\parallel y$}  
 & \multicolumn{2}{|c|}{$\bm {M} \parallel y$, $\bm {E}\parallel x$ }  & \multicolumn{2}{|c}{$\bm {M} \parallel y$, $\bm {E}\parallel y$} \\
 %& \multicolumn{4}{|c|}{$\bm {E} \parallel \bm {x}$}   
 %& \multicolumn{4}{|c}{$\bm {E}  \parallel \bm {y}$} \\
 %\hline 
 %& \multicolumn{2}{|c|}{$\bm {M} \parallel \bm {x}$}  & \multicolumn{2}{|c|}{$\bm {M} \parallel \bm {y}$}  
 %& \multicolumn{2}{|c|}{$\bm {M} \parallel \bm {x}$ }  & \multicolumn{2}{|c}{$\bm {M} \parallel \bm {y}$} \\
\hline 
\hfil  $\hat{O}$ & \hfil $\sigma_v$ & \hfil  $\sigma_h\mathcal{T}$ & $\hfil \sigma_v$ & \hfil $\sigma_h\mathcal{T} $
                        & \hfil $\sigma_v\mathcal{T}$ & \hfil  $\sigma_h\mathcal{T}$ & $\hfil \sigma_v\mathcal{T}$ & \hfil $\sigma_h\mathcal{T} $         \\
\hline 
$ \hfil \bm E \hfil $           & \hfil$-$\hfil & \hfil$+$\hfil & \hfil$+$\hfil & \hfil$+$\hfil & \hfil$-$\hfil & \hfil$+$\hfil & \hfil$+$\hfil & \hfil$+$\hfil \\
\hline 
$ \hfil \tau \hfil $              & \hfil$+$\hfil & \hfil$-$\hfil & \hfil$+$\hfil & \hfil$-$\hfil & \hfil$-$\hfil & \hfil$-$\hfil & \hfil$-$\hfil & \hfil$-$\hfil \\
\hline 
$ \hfil \delta \bm s \hfil $  & \hfil$-$\hfil & \hfil$-$\hfil & \hfil$+$\hfil & \hfil$-$\hfil & \hfil$+$\hfil & \hfil$-$\hfil & \hfil$-$\hfil & \hfil$-$\hfil \\
\hline 
 & \multicolumn{2}{|c|}{ $\delta \bm s \parallel  z$}  & \multicolumn{2}{|c|}{ $\delta \bm s = 0$} 
 & \multicolumn{2}{|c|}{ $\delta \bm s \parallel  z$}  & \multicolumn{2}{|c}{ $\delta \bm s = 0$} \\
\hline 
\end{tabular}
}}
\end{center}
\label{table:1}
\end{table}

 \textcolor{forestgreen}{\emph{\textsf{Calculation results}.}}--- Armed with the above symmetry analysis, we calculate the electronic properties of VSe$_2$ monolayer by first-principles and Wannier function approaches, in order to study the exotic CISP and possible valley-related physics.  The calculation method can be found in Supplemental Material \citep{SM}. %Considering spin exchange interaction in the calculations, we first confirm that the ferromagnetic ordering is energetically favoured, while the antiferromagnetic and non-magnetic configurations is respectively XX meV and XX meV higher in energy. 
The calculated lattice constant of VSe$_2$ monolayer is 3.33  \AA\  and the stable ferromagnetic order has a magnetic moment of 1.0 $\mu_B$ per unit cell, in agreement with previous results \citep{Fuh16}.  Given that vanadium is one less valence electron than molybdenum, the magnetic moment is mainly contributed by vanadium, compared with non-magnetic MoX$_2$. %Taking into account spin-orbit coupling, the calculated magnetiocrystalline anisotropy energies (MAE) indicate an in-plane easy axis of magnetization along $X$ directions. The magnetization along $y$ and $z$ directions give compared with $x$-axis magnetization. 

Fig. \ref{fig:1}(c) and (d) show band structures of VSe$_2$ monolayer with the magnetization along $x$ and $z$ directions, respectively. When the magnetization is along $x$ axis, there are two inequivalent but degenerate valleys at K$_+$ and K$_-$ points in the momentum space. It is also found that the degeneracy is preserved with an in-plane rotation of magnetization. For out-of-plane magnetization, the valley degeneracy is lifted with a splitting of 78 meV by comparing the valence band maxima at the two valleys, similar to the case of magnetized MoX$_2$ \citep{Qi15, Chen17}. The valley splitting leads to novel valley-spin physics. For example, valley polarization can be achieved by both carrier doping and non-polarized optical pumping, besides well-known chiral optical field \citep{Qi15, Tong16, Chen17}. The valley structure and its opto-electronic responses are therefore tunable by tilting the magnetization. The out-of-plane CISP, if realized, provides opportunities for electric control of the magnetization and further electric control of valley properties in VSe$_2$ monolayer. On the other hand, the valley degree of freedom may add a new dimension to CISP, if CISP has valley dependence.

\begin{figure}[h!]
\centering
\includegraphics[width=8.5cm]{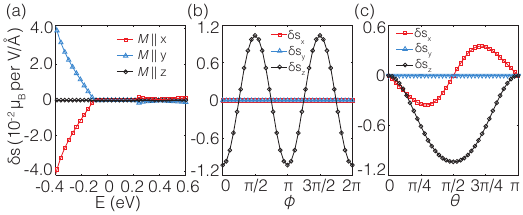}
\caption{The calculated $\mathcal{\delta} \bm s$ in a unit cell, under the electric field along $x$ axis. (a) $\mathcal{\delta} \bm s_z$  as a function of Fermi energy. The valence band maximum is set to zero energy. Red, blue and black lines respectively correspond to the magnetization along $x$, $y$, $z$ direction.  (b) and (c) The evolution of $\mathcal{\delta} \bm s$ with in-plane rotation of magnetization and  out-of-plane rotation within the $x$-$z$ plane. $\phi=0$ and $\theta=0$ respectively correspond to $x$ and $z$ directions. The Fermi level is set to  -0.2 eV.}
\label{fig:2}
\end{figure}

We further compute CISP in VSe$_2$ monolayer by the Kubo linear response formula, where the coefficient $\chi$ is given as \cite{Kurebayashi14},
 \begin{equation}
\chi  = \frac{e \Gamma}{\pi  S} \text {Re}  \sum_{n,\bm {k}} \langle \bm {s}\rangle_{n\bm {k}}  \langle \bm {v}\rangle_{n\bm {k}} 
(G_{n\bm {k}}^{A} G_{n\bm {k}}^{R}-G_{n\bm {k}}^{R} G_{n\bm {k}}^{R}) .
\label{H}
\end{equation}
Here, each electronic state is denoted by the band index $n$ and the momentum $\bm k$. $\langle \bm {s}\rangle_{n\bm {k}} $ and $\langle \bm {v}\rangle_{n\bm {k}} $ are the expectation values of the equilibrium spin and the velocity, respectively.  The velocity operator $\bm {v}=\frac{1}{\hbar}\frac{\partial H}{ \partial {\bm k}}$, of which the direction is determined by $\bm E$.  $G_{n\bm {k}}^{R}$ and $G_{n\bm {k}}^{A}$ are respectively the retarded and advanced Green functions, defined as $G_{n\bm {k}}^{R}=(G_{n\bm {k}}^{A})^*=1/(E_F-E_{n\bm {k}}+i\Gamma)$. $E_F$ and $E_{n\bm {k}}$ are respectively the Fermi and  electronic energies. $\Gamma$ is the band broadening due to the finite lifetime of electronic states in the presence of disorder, which is related to the relaxation time by $\Gamma=\hbar/2\tau$. $\Gamma$ is set to 0.01 eV, by reference to another MX$_2$ \citep{Radisavljevic11, Zhang14}. %carrier mobility of ~10$^2$ cm$^2$V$^{-1}$s${^{-1}}$ and an effective mass comparable to electron mass
 $e$ is the electron charge and $S$ is the monolayer's area. %$\hbar$ is the reduced Planck constant. 

Fig. \ref{fig:2}(a) show the calculated CISP in VSe$_2$ monolayer with the shift of the Fermi level, under the electric field along $x$ axis. There are indeed non-vanishing vertical CISPs for the magnetization along both $x$ and $y$ directions, while the in-plane component is zero. The two in-plane magnetizations give opposite CISPs, indicating an in-plane  anisotropy. The CISP of the valence band states are larger than those of the conduction band in magnitude and it has an order of 10$^{-7} \mu_B$ per A/cm, when normalized by a longitudinal conductivity \citep{SM}.  With the monolayer
thickness of about 6 \AA\ considered \citep{Li14d}, this corresponds to an order of 10$^{-8} \mu_B$ per 10$^{7}$ A/cm$^2$, comparable to CISP in Mn$_2$Au \cite{Zelezny17}. Given that considerable doping level \citep{Brumme15} and current density \citep{Lembke12} have been achieved in MX$_2$, the CISP and associated torque in VSe$_2$ are expected to be readily observable experimentally. For the vertical  magnetization, $\mathcal{\delta} \bm s=0$ due to the $C_3$ symmetry. Under the electric field along $y$ axis, the corresponding $\mathcal{\delta} \bm s$ is also vanishing. These results agree well with symmetry analysis.

The CISP as a function of the magnetization direction is further shown in Fig. \ref{fig:2} (b) and (c), with the direction denoted by the elevation angle $\theta$ and the azimuth angle $\phi$ . For in-plane magnetization ($\theta=\pi/2$),  $\mathcal{\delta} \bm s_z$ is proportional to cos2$\phi$. The second-order directional dependence may be exploited to reverse the CISP.  For out-of-plane magnetization within the $x$-$z$ plane ($\phi=0$), $\mathcal{\delta} \bm s_z$ always points to the same direction except for vanishing $\mathcal{\delta} \bm s_z$ at $\theta=0$ and $\pi$. Besides,  $\mathcal{\delta} \bm s_x$ appears as well. % and the ratio of $\mathcal{\delta} \bm s_z$/$\mathcal{\delta} \bm s_x$ satisfies a relation of 1.47 tan$\theta$.
% The intra-band term give a field-like contribution pointing to a certain direction, act like a magnetic field. 

\begin{figure}[h!]
\centering
\includegraphics[width=8cm]{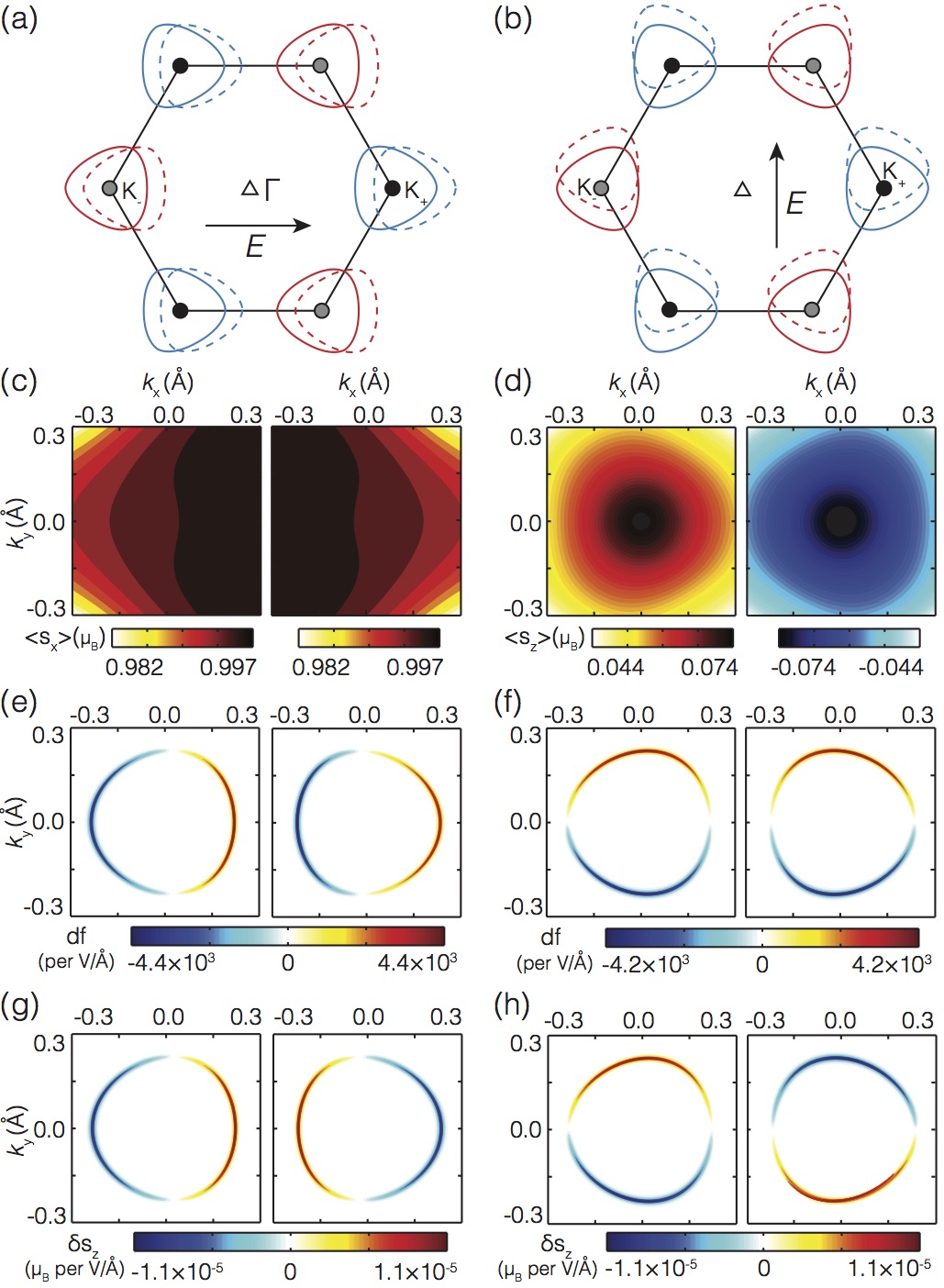}
\caption{Momentum-resolved electronic structures near $K_{\pm}$ valleys,  with a magnetization along $x$ direction. (a) and (b) Schematics of the carrier redistribution under the electric field along $x$ and $y$ directions. (c) and (d) the $x$ and $z$ components of the equilibrium spin.  (e) and (f) the change of distribution function for two electric field directions. (g) and (h) $\mathcal{\delta} \bm s_z$ for two electric field directions. In (c)-(h), left and right panels respectively correspond to $K_{-}$ and  $K_{+}$ valleys. The Fermi level is set to -0.2 eV. }
\label{fig:3}
\end{figure}

To study the contribution to CISP from $K_{\pm}$ valleys, Fig. \ref{fig:3} shows momentum-resolved electronic structures in the neighborhood of the two valleys, with a magnetization along $x$ axis considered. Fig. \ref{fig:3} (a) and (b) schematically illustrate the redistribution of carriers on the Fermi level. The valley pockets are shifted along electric field directions.  Based on this picture, $\mathcal{\delta} \bm s$ can be also described as  $\mathcal{\delta} \bm s=\frac{1}{S}\sum_{n\bm {k}} \langle \bm {s}\rangle_{n\bm {k}}\delta f_{n\bm {k}} $ by the semiclassical Boltzmann approach, besides the Eq. \ref{ef}. The change of the distribution function, 
 $\delta f_{n\bm {k}}=\frac{eE \hbar}{2\pi } \langle \bm {v}\rangle_{n\bm {k}}  \text {Re}(G_{n\bm {k}}^{A} G_{n\bm {k}}^{R}-G_{n\bm {k}}^{R} G_{n\bm {k}}^{R})$, according to Eqs. \ref{ef} and \ref{H}. Taking into account the highest valence band that crosses the chosen Fermi level, $\langle \bm {s}\rangle_{\bm {k}}$, $\delta f_{\bm {k}}$ and $\mathcal{\delta} \bm s_{\bm {k}}$ are given in Fig. \ref{fig:3} (c)-(h).  Since the two valleys are related by $\sigma_v$, %i.e. the $k_y=0$ plane,
  these quantities in two valleys are symmetric or antisymmetric with respect to $\sigma_v$.  For the spin expectation,  its component along the magnetization direction, $\langle \bm {s_x}\rangle$, is dominant and has the same sign in two valleys.  $\langle \bm {s_z}\rangle$ appears as well, which arises from intrinsic spin-orbit coupling in the presence of $\sigma_h$ \citep{Zhu11,Xiao12}. $\langle \bm {s_z}\rangle$ is much smaller than $\langle \bm {s_x}\rangle$, indicating that the spin-orbit coupling is much smaller than the exchange interaction.  More importantly, $\langle \bm {s_z}\rangle$ is opposite in two valleys, because of the valley-distinct Zeeman-type spin splittings induced by intrinsic spin-orbit coupling \citep{Zhu11,Xiao12, Tong16,SM}, in contrast with $\langle \bm {s_x}\rangle$.  On the other hand,  $\delta f$ in two valleys has distinct responses to the electric field directions. When $\bm E \parallel x$, a mirror-antisymmetric $\delta f$ demonstrates a current-induced valley polarization, that is, a carrier transfer between two valleys. Combining valley-contrasting $\langle \bm {s_z}\rangle$, two valleys equally contribute to a non-vanishing CISP.  In contrast, there is no carrier valley polarization and CISP when $\bm E \parallel y$. %Besides, when the magnetization tilts from two-dimensional plane, the valley splitting gives rise to valley-dependent contributions to CISP even if $\bm E \parallel x$ \cite{SM}. 
  Therefore, the valley polarization, together with valley-contrasting $\langle \bm {s_z}\rangle$, is fundamental to vertical CISP. A rotated in-plane electric field proposed here paves a highly-controllable way for  valley polarization. Besides, valley-polarized current can selectively be introduced by various valley filters \cite{Gunlycke11,Qi15, Tong16, Tu17}, where the currents with different valley indices can lead to opposite $\mathcal{\delta} \bm s_z$ (See Fig. 3 (d)).

 %That is, it is mirror-antisymmetric and mirror-symmetric when $\bm E \parallel x$ and $\bm E \parallel y$, respectively. By combining valley-distinct $\langle \bm {s_z}\rangle$ and  $\bm E$-dependent $\delta f$ Therefore, two valleys can make the same or opposite contributions, depending on the direction of electric field. Besides, when the magnetization tilts from two-dimensional plane, the valley degeneracy is lifted, which gives rise to distinct contributions from two valleys even if  $\bm E \parallel x$ \cite{SM}.
 %Though small, $\langle \bm {s_z}\rangle$'s valley distinction, associated with mirror-antisymmetric $\delta f$, leads to the same contribution of non-vanishing $\mathcal{\delta} \bm s_z$ from two valleys when $\bm E \parallel x$. In contrast, the contributions from two valleys cancel each other when $\bm E \parallel y$, by combining valley-distinct $\langle \bm {s_z}\rangle$ and mirror-symmetric $\delta f$. 

 % The point group changes from $D_{3h}$ to $C_{2v}$ under the uniaxial strain.

\begin{figure}[h!]
\centering
\includegraphics[width=8.5cm]{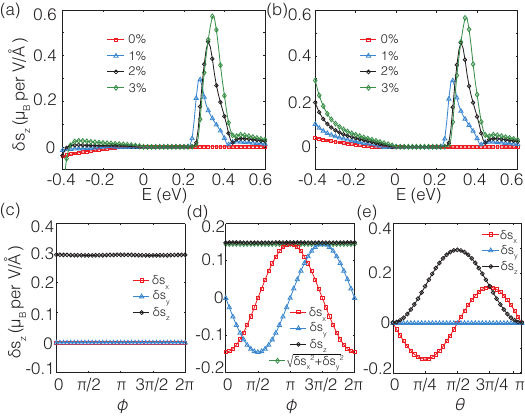}
\caption{$\mathcal{\delta} \bm s$ under a uniaxial strain. (a) and (b) $\mathcal{\delta} \bm s_z$  as a function of Fermi energy, with the magnetization along $x$ and $y$ directions.  (c)-(e)  $\mathcal{\delta} \bm s$ as rotating the magnetization within the $\theta=\pi/2$, $\theta=\pi/4$ and $\phi=0$ planes, repectively. A 1\% strain and $E_F$=0.28 eV where a peak is located in (a) and (b) are adopted. }
\label{fig:4}
\end{figure}

We further compute CISP in VSe$_2$ monolayer under a uniaxial tensile strain along $y$ axis, as shown in Fig. \ref{fig:4}.  The crystal point group is reduced to $C_{2v}$, with the principal axis along the $y$ direction. The $C_3$ symmetry is broken structurally, while both $\sigma_v$ and $\sigma_h$ are kept, which still satisfy symmetry requirements for vertical CISP. The two valleys are well preserved, according to the band structure of the strained monolayer in Fig. \ref{fig:1} (c) and (d).
For the CISP, $\mathcal{\delta} \bm s_z$ is still only non-vanishing component with in-plane magnetization. It has increased maxima, when turning away from the $C_3$ symmetry. Comparing the magnetization along $x$ and $y$ directions, $\mathcal{\delta} \bm s_z$ becomes isodirectional near the band edges (Fig. \ref{fig:4} (a) and (b)).  Especially for the conduction bands, the enhanced $\mathcal{\delta} \bm s_z$ has almost no azimuth-angle-dependence, demonstrating a good isotropy (Fig. \ref{fig:4} (c)). When the magnetization is tilted from the two-dimensional plane,  $\mathcal{\delta} \bm s_z$ keeps isotropic for another latitude with a certain $\theta$, as well as emergent in-plane components (Fig. \ref{fig:4} (d)). The isotropic $\mathcal{\delta}\bm s_z$ always exists including small quantities at $\theta=0/\pi$ and points to +$z$ axis (Fig. \ref{fig:4} (e)). Similar results are also found under the compressive strain \cite{SM}.  The in-plane directional dependence and magnitude of the CISP are therefore tuned by a uniaxial strain. %The $\mathcal{\delta}\bm s_z$ with the strain is therefore field-like and may give rise to a perpendicular magnetization reorientation. 
 
%Maximum experimental charge-carrier concentration n per area cm$^{−2}$ for the different dichalcogenides using either a solid-state or an ionic-liquid-based FET \citep{Brumme15}.

We also calculate the induced spin polarization from Berry phase contribution \citep{Zelezny14, Zelezny17, Xiong17}. It is one order of magnitude smaller than that from the carrier redistribution. Besides, the induced spin is in-plane, in contrast with the above CISP.

 \textcolor{forestgreen}{\emph{\textsf{Summary}.}}---We demonstrate that out-of-plane spin polarization emerges in VSe$_2$ monolayer under electric current. The vertical non-equilibrium spin arises from the intrinsic spin-orbit coupling, distinct from widely discussed Rashba systems. By introducing electric field direction, valley currents and strain field, the induced spin can be tuned in controllable ways, which has potential use in perpendicular magnetization switching and electric control of valleys. 
 
 It is noted that VSe$_2$ monolayer is an initial demonstration of the vertical CISP. There are a number of two-dimensional ferromagnetic systems that have similar symmetries and possible out-of-plane CISP, such as  VS$_2$ \citep{Fuh16}, NbX$_2$ \citep{Xu14}, TaX$_2$ \citep{Manchanda15},  group-III monochalcogenides \cite{Cao15} and so on.  On the other hand, even if non-magnetic, the uniaxial strained MX$_2$ and another monolayer with the $C_{2v}$ symmetry may also exhibit out-of-plane CISP, based on the above symmetry analysis. Moreover, there are both intrinsic and Rashba spin-orbit couplings in the heterostructure of MX$_2$/ferromagnet \cite{Shao16, Zhang16, Zhao17}. While intrinsic spin-orbit coupling was previously ignored, their cooperative roles in generating CISP are worth further study. The intrinsic spin-orbit coupling may be also exploited to create non-equilibrium spin in collinear and non-collinear antiferromagnetic systems \citep{Zelezny14, Wadley15, Zelezny17, Li13, Chen14}

%Maximum experimental charge-carrier concentration \cite{Brumme15}. The mobility \cite{} The valley current is introduced give distinct CISP.
%An important topic of spintronics research is how charge and spin currents may affect the orientation and dynamics of magnetic moments in a magnetized material including both ferromagnets and antiferromagnets. There are potentially two-dimensional materials with spontaneous magnetic ordering, and the order parameter has strong coupling to the electronic band structure.  For example, proximity magnetic coupling of MX2 (M=Mo,W and X=S Se, Te) to a ferromagnetic insulator (EuO)  lifts valley degeneracy with a giant energy splitting .  MnPX3 (X=S, Se), has an antiferromagnetic spin ordering in a honeycomb lattice, which directly couples to the valley degrees of freedom.  We expect that charge and valley currents in two-dimensional materials may strongly affect the orientation and dynamics of the magnetic moments in the material or substrate, which may be applied for magnetization switching and domain-wall motion in memory/logic devices .
%   We have already done preliminary calculations on 

%We will extend such calculations to more realistic two-dimensional materials with ferromagnetic/antiferromagnetic order. We will also keep an eye on oxides, which show a similar strong dependence of band structure, valley splitting, and Berry curvature on the direction of magnetization ordering 

\bigskip

\begin{acknowledgments}
X. L. is grateful to Shiang Fang, Jakub {\v{Z}}elezn{\`y} and Huawei Gao for valuable discussions.
This work is supported by  DOE (DE-FG03-02ER45958, Division of Materials Science and Engineering).
\end{acknowledgments}

%\bibliography{references}

\end{document}